\documentclass[traditabstract]{aa} 
\UseRawInputEncoding
\usepackage{amsmath}
\usepackage{graphicx}
\usepackage{txfonts}
\usepackage{natbib,twoopt}
\usepackage[breaklinks=true]{hyperref} 
\bibpunct{(}{)}{;}{a}{}{,} 
\makeatletter
\newcommandtwoopt{\citeads}[3][][]{\href{http://adsabs.harvard.edu/abs/#3}%
{\def\hyper@linkstart##1##2{}%
\let\hyper@linkend\@empty\citealp[#1][#2]{#3}}}
\newcommandtwoopt{\citepads}[3][][]{\href{http://adsabs.harvard.edu/abs/#3}%
{\def\hyper@linkstart##1##2{}%
\let\hyper@linkend\@empty\citep[#1][#2]{#3}}}
\newcommandtwoopt{\citetads}[3][][]{\href{http://adsabs.harvard.edu/abs/#3}%
{\def\hyper@linkstart##1##2{}%
\let\hyper@linkend\@empty\citet[#1][#2]{#3}}}
\newcommandtwoopt{\citeyearads}[3][][]%
{\href{http://adsabs.harvard.edu/abs/#3}
{\def\hyper@linkstart##1##2{}%
\let\hyper@linkend\@empty\citeyear[#1][#2]{#3}}}
\makeatother
\begin{document}

   \title{On the X-ray pulsar HD 49798: a contracting white dwarf with debris disk?}

   \titlerunning{X-ray pulsar HD 49798:a contracting WD with debris disk?}

   \authorrunning{Wen-Cong Chen }

   \author{Wen-Cong Chen\inst{1,2}}

    \institute{School of Science, Qingdao University of Technology, Qingdao 266525, China
    \and
              School of Physics and Electrical Information, Shangqiu Normal University, Shangqiu, Henan 476000, China
        \\chenwc@pku.edu.cn
             }
   \date{}

\abstract{HD49798/RX J0648.0¨C4418 is a peculiar binary including a hot subdwarf of O spectral type and a compact companion in a 1.55 day orbit. According to the steady spin period derivative $\dot{P}=(-2.17\pm0.01)\times10^{-15} ~\rm s\,s^{-1}$ , the compact object was thought to be a contracting young white dwarf (WD). However, the X-ray luminosity producing by the wind accretion of massive WD is one order of magnitude smaller than the observed value. In this work, we propose an alternative model to account for the observed X-ray luminosity. If the WD was surrounded by a debris disk, the accretion from the debris disk can produce the observed X-ray luminosity and X-ray pulses. Based on the time-varying accretion rate model, the current mass of the debris disk is constrained to be $3.9\times10^{-6}~\rm M_{\odot}$. Comparing with the contraction of the WD, the accretion torque exerting by such a debris disk can only influence the spin evolution of the WD in the early stage. According to the accretion theory, the magnetic field of the WD is constrained to be $\sim (0.7-7)\times10^{4}$ G. The calculated conventional polar cap radius of the WD is larger than the observed emitting-zone radius, which probably originate from the existence of strong and small-scale local magnetic field in the polar cap surface. We expect that further multiband observations on this source can help us to confirm or rule out the existence of a debris disk.}

\maketitle

\section{Introduction}

HD 49798/RX J0648.0$-$4418 is a peculiar binary including a hot subdwarf of O spectral type and a compact companion in an orbit with an orbital period $P_{\rm orb}=1.55$ day \citep{thac70,kudr78}. When this source was discovered, it was the brightest hot subdwarf detected \citep{jasc63}, and is still one of the brightest hot subdwarf so far \citep{mere11}. Bisscheroux et al. (1997) suggested that an intermediate-mass star that entered into a common envelope while on the early AGB stage is the most likely progenitor of HD 49798.

Israel et al. (1995, 1997) had detected a 13.2 s period X-ray pulse, which probably originated from the spin period ($P$) of a magnetic compact object accreting from the weak wind of subdwarf, in which the wind loss rate is about $3\times10^{-9}~\rm M_{\odot}\,yr^{-1}$ \citep{hama10}. XMM-Newton data from 2002 to 2014 derived a relatively low X-ray luminosity $L_{\rm X}\approx (1.3\pm0.3)\times10^{32}(d/520\rm pc)^{2}~erg\,s^{-1}$ \citep[$d$ is the distance of the source,][]{mere16}. Comparing the observed X-ray luminosity with the accretion luminosity estimated by the wind capture rate of the compact object, Israel et al. (1996) proposed that the X-ray pulsator should be a neutron star (NS) rather than a white dwarf (WD). However, a very soft blackbody of temperature ($kT\sim30~\rm eV$), hard power-law tail, and large emitting area radius ($R_{\rm BB}\sim32(d/520\rm pc)~\rm km$) derived from the blackbody spectral fit tended to a WD compact object \citep{mere09,mere11}. Most recently, a relatively precise parallax obtained with Gaia EDR3 measured the distance of this source to be $521\pm14~\rm pc$ \citep{brow20}.

Based on the data from \emph{XMM-Newton} satellite, Mereghetti et al. (2009) obtained an X-ray mass function, and an orbital plane inclination angle ($79^{\circ}$ to $84^{\circ}$) by detecting an eclipse in the X-ray light curve, and constrained the mass of X-ray pulsator to be $1.28\pm0.05~\rm M_{\odot}$, the mass of the hot subdwarf to be $1.50\pm0.05~\rm M_{\odot}$. Adopting the optically thick wind assumption, Wang \& Han (2010) proposed that HD 49798/RX J0648.0$¨C$4418 could produce a type Ia supernova by the accretion of CO WD in the future. Recently, \cite{liu15} argued that the X-ray pulsar companion of HD 49798 is a CO WD rather than a ONe WD by the binary population synthesis simulation. If HD 49798 accompanied by a NS, this source will appear as an ultraluminous X-ray source by the mass transfer triggered by Roche lobe overflow in the future, and eventually evolve into a wide intermediate-mass binary pulsar \citep{broo17}. \cite{wu19} found that the WD would experience an off center carbon burning and form a neutron star via Fe core collapse supernova if the compact companion of HD 49798 is a CO WD. However, this source is unlikely to form a neutron star from an accretion-induced-collapse process if the compact object is a ONe WD \citep{liu18}.

It is still controversial that the X-ray pulsator companion of HD 49798 is NS or WD. Mereghetti et al. (2016) performed a phase-connected timing analysis for \emph{XMM-Newton}, \emph{Swift}, and \emph{ROSAT} data spanning more than 20 yr, and obtained the spin-period derivative of the X-ray pulsator to be $\dot{P}=(-2.15\pm0.05)\times10^{-15} ~\rm s\,s^{-1}$. Recently, the compact companion was reported to be still spinning up at a steady rate of $\dot{P}=(-2.17\pm0.01)\times10^{-15} ~\rm s\,s^{-1}$ according to the new \emph{XMM-Newton} data \citep{mere21}. In principle, an accretion process of compact object can result in a steady spin-up rate. The orbital separation of HD 49798 is about $8.0~R_{\odot}$ \citep{mere09}, the effective Roche-lobe radius of the donor star can be estimated to be $\sim3.1~R_{\odot}$ \citep{eggl83}. The radius of the subdwarf is $1.05\pm0.06~R_{\odot}$ \citep{krti19}, hence it is impossible to transfer material by the Roche-lobe overflow. However, this source is a member of the few hot subdwarfs offering obvious evidence for a stellar wind \citep{hama81,hama10}. The observed spin-up rate favors a NS accreting from the wind of the hot subdwarf. However, if the compact companion of HD 49798 is a NS, there still exist three puzzles in the NS model: first, the inferred low magnetic field ($\sim10^{10}$ G) is unusual for a NS without millisecond period \citep{mere16}; second, it is impossible to obtain such a large emitting area ($R_{\rm BB}\sim32~\rm km$) fitted by the blackbody spectral; third, such a steady spin-up rate is difficult to interpret for a NS accreting from the stellar winds, in which variations of wind accretion rate captured by the NS would cause the changes of accretion torque, and influence its spin period \citep{mere21}.

If the compact companion of HD 49798 is a WD, it is very difficult to produce the observed spin-up rate by the stellar wind accretion, unless the disk accretion occurs \citep{mere16}. Population synthesis simulations on hot subdwarf binaries also shown that the number of the systems hosting WDs much more than those with NSs \citep{yung05,wu18}. Recently, Popov et al. (2018) provided a novel model, in which the contraction of a young WD with an age of $\sim2~\rm Myr$ can successfully explain the observed spin-up rate. However, the wind accretion model predicated an X-ray luminosity of $L_{\rm X}=1.3\times10^{31}~\rm erg\,s^{-1}$ \citep{krti19} \footnote{It is worth emphasizing that the wind accretion rate strongly depends on the wind velocity at the vicinity of the accreting WD \citep{krti19}. Actually, the effect of ionization by the X-ray flux might decrease the wind velocity, thus increasing the accretion rate \citep{sand18,krti18}.}, which is one order of magnitude smaller than the observed value $L_{\rm X,obs}\approx (1.3\pm0.3)\times10^{32}~\rm erg\,s^{-1}$ \citep{mere16}. Therefore, It still remains a puzzle for this peculiar X-ray source. In this work, we propose an alternative model invoking a debris disk to account for the observed X-ray luminosity.

\section{Debris disk accretion model}\label{sec:2}
Some hot WDs have been discovered surrounding by a debris disk, e. g. SDSS 1228+1040 \citep{gans06,mans19}, SDSS J1043+0855 \citep{gans07}, SDSS J0845+2257 \citep{wils15}, SDSS J1344+0324 \citep{li17}. Recently, the WD G29-38 was reported to be currently accreting planetary material from a debris disk according to the X-ray observations of \emph{Chandra X-ray Observatory} \citep{cunn22}. Although the compact companion of HD 49798 is in a binary system, we assume that it was surrounded by a debris disk similar to these isolated WDs.

The magnetic accretion process of the WD from a debris disk is tightly related to the magnetospheric
(Alfv\'{e}n) radius $r_{\rm m}$, at which the ram pressure of the infalling material is balanced with the magnetic pressure. The magnetospheric radius can be written as \citep{davi73}
\begin{equation}
r_{\rm m}=\xi\left(\frac{\mu^{4}}{2GM\dot{M}^{2}}\right)^{1/7},
\end{equation}
where $G$ is the gravitational constant, $\xi$ a dimensionless parameter of the order unity, $M$ the WD mass, $\dot{M}$ the mass inflow rate at $r_{\rm m}$ in the debris disk, $\mu=B_{\rm p}R^{3}/2$ ($B_{\rm p}$ is the surface dipole magnetic field) the dipolar magnetic momentum of the WD.

Taking $\xi=0.52$ for the disk accretion case \citep{ghos79}, and inserting some typical parameters in equation (7), thus
\begin{equation}
r_{\rm m}=1.2\times 10^{9} \dot{M}^{-2/7}_{14}M^{-1/7}_{1.28}\mu^{4/7}_{30}~\rm cm,
\end{equation}
where $\dot{M}_{14}$ in units of $10^{14}~\rm g\,s^{-1}$, $M_{1.28}$ in units of $1.28~\rm M_{\odot}$, and $\mu_{30}$ in units of $10^{30}~\rm G\,cm^{3}$.

The observed X-ray luminosity of HD 49798 is $L_{\rm X}\approx (1.3\pm0.3)\times10^{32}(d/520\rm pc)^{2}~erg\,s^{-1}$ \citep{mere16}. Ignoring the X-ray luminosity producing by the stellar winds accretion, the accretion rate (i.e. the mass inflow rate at the inner edge of the debris disk) of the accreting WD in HD 49798 can be estimated to be
\begin{equation}
\dot{M}=\frac{L_{\rm X}R}{GM}\approx(2.3\pm0.5)\times 10^{14}~\rm g\,s^{-1},
\end{equation}
where $R$ is the WD radius. In this work, we take $R=3000~\rm km$.

Whether could a debris disk around the WD provide such an accretion rate? After a debris disk forms, the accretion rate should decrease self-similarly in accordance with $\dot{M}\propto t^{-\alpha}$ due to the influence of viscous processes \citep{cann90}. In our debris disk model, an evolutionary law of the accretion rate same to \cite{chat00} is adopted as follows
\begin{equation}
\dot{M}= \left\{\begin{array}{l@{\quad}l} \dot{M}_{0},
& t<T \strut\\
\dot{M}_{0}(t/T)^{-\alpha}
, & t\geq T, \strut\\\end{array}\right.
\end{equation}
where $T$ is of order the dynamical timescale in the inner regions of the debris disk, and $\dot{M}_{0}$ is a constant accretion rate. The initial mass of the disk can be written as
\begin{equation}
M_{\rm d,i}=\dot{M}_{0}T+\int_{T}^{\infty}\dot{M}_{0}(t/T)^{-\alpha}{\rm d}t.
\end{equation}
Therefore, we have $\dot{M}_{0}=(\alpha-1)M_{\rm d,i}/(\alpha T)$ if $\alpha>1$ \citep{chat00}. The dynamical timescale in the inner regions of the debris disk is given by
\begin{equation}
\tau_{\rm dyn}\sim\sqrt{\frac{r_{\rm m}^{3}}{GM}}\sim 2\left(\frac{r_{\rm m}}{10^{9}~\rm cm}\right)^{3/2}\left(\frac{1.28~M_{\odot}}{M}\right)^{1/2}~\rm s.
\end{equation}
In the following calculations, we take $T=1~\rm s$, and $\alpha=19/16$ that the opacity is
dominated by electron scattering \citep{cann90}.

To account for the observed spin-up rate, Popov et al. (2018) proposed that the compact companion of HD 49798 is a contracting WD with a cooling age of $\sim2~\rm Myr$. Similar to G29-38, we also assume that RX J0648.0 - 4418 experienced an accretion for 10\% of the cooling age \citep{jura03b}, i.e. the age of the debris disk is $t_{0}=2\times10^{5}~\rm yr$. To explain the observed X-ray luminosity, the accretion rate from the debris disk should be $\dot{M}=2.3\times10^{14}~\rm g\,s^{-1}$ when $t=t_{0}=2\times10^{5}~\rm yr$, hence the evolution of the accretion rate when $t\geq T$ satisfies
\begin{equation}
\dot{M}=2.3\times10^{14}\left(\frac{t}{2\times10^{5}~\rm yr}\right)^{-19/16}~\rm g\,s^{-1}.
\end{equation}
It yields $\dot{M}_{0}=3.6\times10^{29}~\rm g\,s^{-1}$ from equations (4) and (7), and the initial mass of the debris disk is estimated to be $M_{\rm d,i}\approx0.001~\rm M_{\odot}$ according to equation (5). The current mass of the debris disk is derived to be
\begin{equation}
M_{\rm d,c}=\int_{t_{0}}^{\infty}\dot{M}_{0}(t/T)^{-19/16}{\rm d}t\approx3.9\times10^{-6}~\rm M_{\odot}.
\end{equation}
If the gas-to-dust ratio is 100 \citep{jura03a}, the current dust mass of the debris disk is $M_{\rm dust}\approx3.9\times10^{-8}~\rm M_{\odot}$.

For reference, we also calculate the debris-disk mass of G29-38 using this model. The current accretion rate of G29-38 is about $1.63\times10^{9}~\rm g\,s^{-1}$ \citep{cunn22}, and it has been actively accreting for $4\times10^{7}~\rm yr$, i.e. $t_{0}\approx40~\rm Myr$ \citep{jura03b}. Therefore, the accretion rate from the debris disk is $\dot{M}=1.63\times10^{9}\left(t/40~\rm Myr\right)^{-19/16}~\rm g\,s^{-1}$. Similar to equation (8), the current debris-disk mass can be estimated to be $5.5\times10^{-9}~\rm M_{\odot}$, which is not in contradiction with the minimum disk mass of $\sim10^{-10}~\rm M_{\odot}$ estimated by \cite{jura03b}. Based on three-dimensional radiation-hydrodynamics stellar-atmosphere
models, \cite{cunn21} roughly estimated the debris-disk lifetimes around WDs to be ${\rm log}(t/\rm yr)=6.1\pm1.4$. Therefore, the debris-disk age of G29-38 is probably overestimated. If so, the current disk mass predicted by our model will correspondingly decline.

Taking $M=1.28~\rm M_{\odot}$, $R=3000~{\rm km}$, the X-ray luminosity of the WD accreting from the debris disk can be derived from $L_{\rm X}=GM\dot{M}/R$ and equation (7). Figure 1 plots the evolution of X-ray luminosity (we ignore the radius change of the WD in the contraction stage). It is clear that a WD accreting from a nascent debris disk can appear as luminous X-ray source ($\sim10^{38}\rm ~erg\,s^{-1}$) with a lifetime up to 2 yr, which is similar to a black hole accreting from a fallback disk as an ultraluminous X-ray source \citep{li03}. However, its maximum X-ray luminosity is difficult to exceed $\sim10^{39}\rm ~erg\,s^{-1}$ due to the Eddington luminosity $L_{\rm Edd}=1.9\times10^{38}\rm ~erg\,s^{-1}$ \citep[for the accreting material is hydrogen,][]{king19}. With the decline of the mass inflow rate, the magnetospheric radius will firstly exceed the corotation radius, and the accreting WD transitions to a low X-ray state that lacks X-ray pulsation \citep{camp16}. Subsequently, the magnetospheric radius will then exceed the light cylinder radius ($R_{\rm lc}=cP/2\pi$), and produce a radio emission \citep{illa75,camp98}. According to the critical luminosity that WDs transitions from accretion to propeller regime determined by \cite{camp18}, the accreting WD of HD 49798 will transition to the propeller phase if the accretion luminosity declines to the limiting luminosity $L_{\rm lim}=0.7^{+0.4}_{-0.3}\times10^{32}~\rm erg\,s^{-1}$, which depends the the dipolar magnetic momentum of the accreting WD\footnote{We take $\mu_{30}=0.8$, see also equation (7) of Campana et al. (2018).}.

\begin{figure}
\centering
\includegraphics[scale=0.34,trim={20 30 0 0},angle=0]{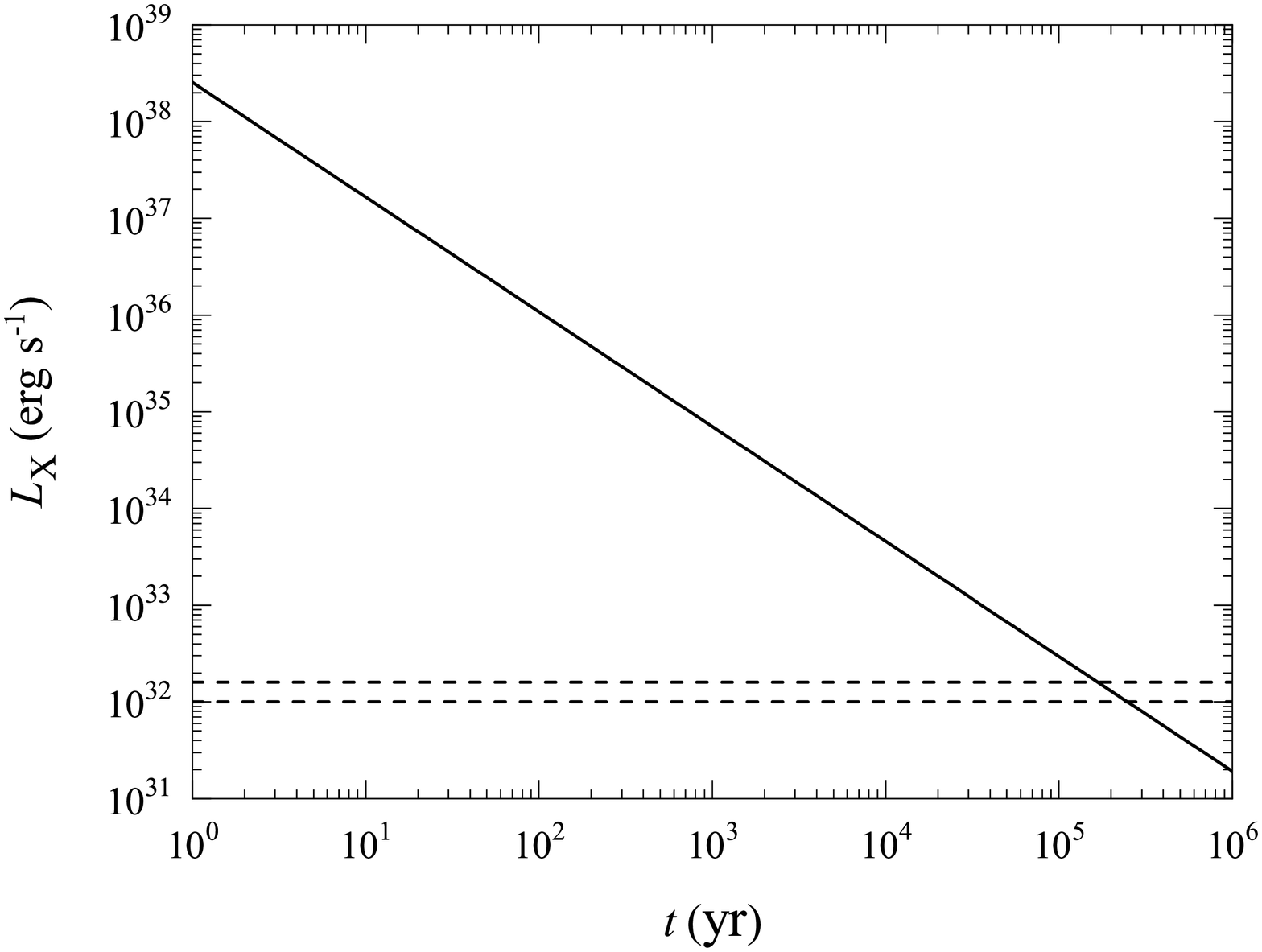}
\caption{X-ray luminosity produced by the accretion from a debris disk as the function of the debris-disk age $t$. We take  $M=1.28~M_{\odot}$, and $R=3000~\rm km$. The horizontal dashed line represent the observed X-ray luminosity range $L_{\rm X}=(1.0-1.6)\times 10^{32}~\rm erg\,s^{-1}$. }
\end{figure}

Similar to the neutron stars, the spin evolution of the WD depends on the interaction
between magnetic field lines and disk plasma, which can give rise to a continuous
exchange of angular momentum between the WD and the disk. If the magnetospheric radius is smaller than the corotation radius (at which the Keplerian angular velocity equals the spin angular velocity of the WD)
\begin{equation}
r_{\rm co} =\sqrt[3]{\frac{GMP^{2}}{4\pi^{2}}}=9.1\times 10^{8}M_{1.28}^{1/3}~\rm cm.
\end{equation}
the WD accretes the specific angular momentum of material at $r_{\rm m}$.

The maximum accretion torque receiving by the WD is $T_{\rm acc}=\dot{M}\sqrt{GMr_{\rm co}}$. Therefore, the maximum
spin-up rate of the WD due to the accretion from a debris disk can be expressed as
\begin{equation}
\dot{P}_{\rm max}=-\frac{P^{2}\dot{M}\sqrt{GMr_{\rm co}}}{2\pi I}\approx-1.1\times10^{-17}P_{13.2}^{2}\dot{M}_{14}M_{1.28}^{2/3}I_{50}^{-1}~\rm s\,s^{-1},
\end{equation}
where $I_{50}$ is the moment of inertia of the WD in units of $10^{50}~\rm g\,cm^{2}$, $P_{13.2}=P/13.2~\rm s$.
For some typical parameters, the maximum spin-up rate producing by an accretion from the debris disk is 1-2 orders of magnitude lower than the observed value. Therefore, the accretion from a debris disk can only account for the observed X-ray luminosity of
HD49798/RX J0648.0-4418.

Figure 2 shows the evolution of the spin-period derivative producing by the accretion from the debris disk. According to our assumption, the debris disk should exist when the WD age is in the range of 1.8 Myr to 2.0 Myr. Comparing with Figure 2 in \cite{popo18}, $\dot{P}$ producing by the debris disk is smaller than that resulting from the WD contraction for the debris-disk age $t=5000-2\times10^{5}~\rm yr$. However, a young (with an age less than 5000 yr) debris disk plays an important role in influencing the spin evolution of the WD.

To support a steady accretion, the inner radius of the debris disk (i.e. the magnetospheric
radius $r_{\rm m}$) should satisfies the following relation as
\begin{equation}
R<r_{\rm m}\leq r_{\rm co}.
\end{equation}
Taking $M_{1.28}=1$, and $\dot{M}_{14}=2.3$, the dipolar magnetic momentum of the WD can be constrained to be
\begin{equation}
0.1<\mu_{30}\leq 0.9.
\end{equation}
Therefore, the surface dipolar magnetic field of the WD is in the range of $(0.7-7)\times 10^{4}~\rm G$.

For a magnetic WD, the accretion flow along the magnetic field lines would form an accretion column inside the polar cap \citep{shap83}. The polar cap opening angle of the last open field line is \citep{rude75}
\begin{equation}
\theta_{\rm open}=\sqrt{\frac{R}{R_{\rm LC}}},
\end{equation}
where $R_{\rm LC}=cP/2\pi$ is the radius of the light cylinder. So we can estimate the polar cap radius of the WD in HD 49798 to be
\begin{equation}
R_{\rm dp}=R\sqrt{\frac{R}{R_{\rm LC}}}=200P_{13.2}^{-1/2}~\rm km.
\end{equation}
This radius is six times as large as the observed radius of the emitting area $R_{\rm BB}\approx32(d/520\rm pc)~\rm km$ at a distance of 520 pc \citep{mere16,brow20}.

Although the estimated polar cap radius is larger than the radius of the emitting zone derived from the blackbody spectral fit, it is already noted that the radius $R_{\rm dp}$ of the conventional polar cap is ten times larger than the one of the radiation area in the neutron star field \citep{herm13,szar17,gepp17}. Strong and small-scale local magnetic field structures in the polar cap surface was thought to be responsible for the small actual radius $R_{\rm pc}$ of polar cap \citep{szar15,szna20}. According to the magnetic flux conservation law, if the magnetic field at the actual polar cap of the WD $B_{\rm s}\simeq36B_{\rm p}=(2.5-25)\times 10^{5}~\rm G$, the small radius of the emitting zone can be easily understood.

\begin{figure}
\centering
\includegraphics[scale=0.34,trim={20 30 0 0},angle=0]{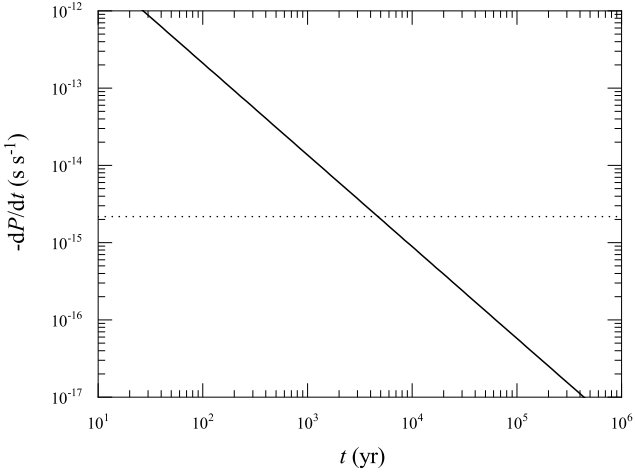}
\caption{Evolution of the spin period derivative of the WD accreting from a debris disk in $\dot{P}$ vs. debris-disk age diagram. We ignore the influence of the spin-period change on the $\dot{P}$. The horizontal dotted line represent the observed spin-up rate $\dot{P}=(-2.17\pm0.01)\times10^{-15} ~\rm s\,s^{-1}$. }
\end{figure}

\section{Summary and discussion}
Stellar wind accretion from the hot subdwarf is insufficient to produce the observed X-ray luminosity of HD 49798 \citep{krti19}. In this work, we propose an alternative model to account for the observed X-ray luminosity of HD 49798. If the compact companion of HD 49798 is a WD surrounding by a debris disk, by the interaction between the magnetic field and the debris disk, the accretion flow along the magnetic field lines would produce an accretion column on the polar cap of the WD, thereby naturally resulting in the observed X-ray pulses. Based on the model of time-varying accretion from a debris disk given by \cite{chat00} and the observed X-ray luminosity, the initial mass and the current mass of the debris disk are constrained to be $\sim0.001~\rm M_{\odot}$ and $3.9\times10^{-6}~\rm M_{\odot}$, respectively. Based on the accretion theory, the surface magnetic field of the WD is constrained to be $B_{\rm p}=(0.7-7)\times 10^{4}~\rm G$, while the small polar cap zone requires a relatively strong local magnetic field ($B_{\rm s}=(2.5-25)\times 10^{5}~\rm G$) to account for a small emitting area. Comparing with the contraction of the WD, the accretion torque exerting by the proposed debris disk can only influence the spin evolution of the WD when the debris-disk age is less than 5000 yr. Therefore, the debris disk cannot spin the accreting WD up to the observed rate in the current stage, which should arise from a change of the moment of inertia of the WD at the contraction stage \citep{popo18}.

The blackbody spectral fit for HD 49798 inferred a radius of the emitting area to be $R_{\rm BB}\approx32(d/520\rm pc)~\rm km$ \citep{mere16}. This emitting area should be the real polar cap zone resulting from the accretion column on the surface of the WD. However, our calculated polar cap radius is 200 km. Difference between the real polar cap zone and the theoretical vale probably originate from strong and small-scale local magnetic field structures in the polar cap surface \citep{szar15,szna20}.

The debris disks around some isolated WDs probably originate from the tidal disruption of either comets \citep{debe02} or asteroids \citep{jura03b}. Our scenario predicts a heavy debris disk with a mass of $\sim 10^{-6}~\rm M_{\odot}$, which is four orders of magnitude higher than that in the WD G29-38 \citep{jura03b}. This mass discrepancy should arise from different origin of debris disks. Since HD 49798 may experienced a common envelope evolutionary phase \citep{biss97}, the debris disk around the WD may originate from the engulfment of the progenitor envelop of the hot subdwarf. For example, the engulfment of a low mass companion star of HD 233517 when it evolved into a red giant results in a heavy debris disk of $\sim0.01~\rm M_{\odot}$ \citep{jura03a}. However, it is challenging to confirm the debris disk by detecting the infrared excess from RX J0648.0 - 4418 like G29-38. First, the distance of RX J0648.0 - 4418 is longer than G29-38 by a factor of 40; second, the detecting radiation flux from the debris disk should be slight due to a large orbital plane inclination angle\footnote{The debris disk should be in a same plane with the orbital plane if it comes from the engulfment of the progenitor envelop of the hot subdwarf in the common envelope stage.} \citep[$79^{\circ}$ to $84^{\circ},$][]{mere09}. In the other hand, it will also confirm the existence of a debris disk if a low X-ray state from RX J0648.0 - 4418 or a spin-down rate are luckily detected in the future. We expect that further multiband observations on this source can help us to confirm or rule out the existence of a debris disk.

\begin{acknowledgements}
We cordially thank the anonymous referee for deep and constructive comments improving this manuscript. This work was partly supported by the National Natural Science Foundation of China (under grant numbers 11573016, 11733009), Natural Science Foundation (under grant number ZR2021MA013) of Shandong Province.
\end{acknowledgements}

\end{document}